\documentclass[twocolumn,preprintnumbers,prl]{revtex4}
\usepackage{graphicx}
\usepackage{times}
\usepackage{epsfig}
\usepackage{amsmath}
\usepackage{amsfonts}
\usepackage{amssymb}
\usepackage{url}
\usepackage{subfigure}
\newcommand{\be}{\begin{equation}}
\newcommand{\ee}{\end{equation}}
\newcommand{\bea}{\begin{eqnarray}}
\newcommand{\eea}{\end{eqnarray}}

\begin{document}
\pagestyle{plain}
\title{
SuperWIMP dark matter and 125 GeV Higgs boson in the minimal GMSB 
}
\author{Nobuchika Okada} 
\affiliation{
Department of Physics and Astronomy,
University of Alabama,
Tuscaloosa, Alabama 35487, USA
}
\begin{abstract}

Recently, both the ATLAS and CMS experiments have observed 
 an excess of events that could be the first evidence 
 for a 125 GeV Higgs boson. 
We investigate an implication of the CP-even Higgs boson 
 with mass around 125 GeV in the context of the minimal 
 gauge mediated supersymmetry breaking (mGMSB). 
In mGMSB, gravitino is the lightest sparticle (LSP)
 and hence the dark matter candidate. 
We consider the so-called superWIMP scenario 
 where the dark matter gravitino is non-thermally produced 
 by the decay of the next-to-LSP (NLSP) bino-like 
 neutralino after its freeze-out. 
For a given $\tan \beta$ and the number of the messengers ($N_m$) fixed, 
 we find that the rest of the mGMSB parameters, 
 the SUSY breaking parameter and the messenger scale, 
 are completely fixed by the conditions of simultaneously 
 realizing the observed dark matter abundance and 
 the 125 GeV Higgs boson mass, leading to 
 the NLSP neutralino mass around $1.5-2$ TeV 
 and the gravitino mass around $3-7$ GeV, 
 depending on the values of $\tan \beta$ and $N_m$. 
The lifetime of the NLSP is found to be shorter than 1 sec, 
 so that the success of the big bang nucleosynthesis remains intact. 
The non-thermally produced gravitino behaves as the warm dark matter  
 with the free-streaming scale found to be 
 $\lambda_{\rm FS} \simeq 0.1$ Mpc, 
 whose value is reasonable for observations of the power spectrum 
 on both large and sub-galactic scales in the Universe. 

\end{abstract}
\maketitle


The low energy supersymmetry (SUSY) is arguably one 
 of the most promising candidates for new physics 
 beyond the Standard Model (SM). 
The minimally supersymmetric extension of the SM (MSSM) 
 provides us with not only a solution to the gauge hierarchy 
 problem but also various interesting phenomena such as 
 the successful SM gauge coupling unification, 
 the radiative electroweak symmetry breaking, 
 prediction of the SM-like Higgs boson mass, 
 and a candidate for the dark matter in our Universe. 
It has been expected that some of sparticles 
 can be discovered in the near future, 
 most likely at the Large Hadron Collider (LHC).

As well as the discovery of new particles, 
 the discovery of the Higgs boson is another major goal 
 of the physics program at the LHC, in order to confirm 
 the origin of the electroweak symmetry breaking and 
 the mechanism of particle mass generation. 
Recently, the ATLAS~\cite{ATLAS} and CMS~\cite{CMS} experiments 
 have reported an excess of evens that could be 
 the first evidence of the Higgs boson with mass 
 of around 125 GeV~\cite{Moriond}. 
The observations are supported by recent analysis of 
 the Tevatron experiments~\cite{Moriond}. 
The 125 GeV Higgs boson has a great impact on SUSY phenomenology, 
 because the MSSM has a prediction of the SM-like Higgs boson mass 
 as a functions of soft SUSY breaking parameters, 
 in particular, stop masses. 
Detailed studies for realizing the Higgs mass around 125 GeV 
 is a current hot topic in SUSY phenomenology~\cite{125Higgs-MSSM}. 
In most of the studies, the constrained MSSM (CMSSM) or 
 slight extention of the CMSSM is adopted for the boundary conditions 
 on the soft SUSY parameters at the scale of 
 the grand unified theory (GUT).

Although the CMSSM or more generally, 
 the context of supergravity mediated SUSY breaking 
 is a very simple benchmark in examining 
 the sparticle mass spectrum, this scenario potentially 
 suffers from the SUSY flavor problem~\cite{RS}. 
The gauge mediated SUSY breaking (GMSB)~\cite{GMSB} offers 
 the compelling resolution for the SUSY flavor problem 
 due to the SUSY breaking mediation via the flavor-blind 
 SM gauge interactions. 
In this paper, we investigate an implication of the 125 GeV Higgs 
 in the context of the mGMSB. 
There were several studies on this context~\cite{GMSB-125a},  
 and very recently a more comprehensive studies with the parameter scan 
 has been performed~\cite{GMSB-125b}. 
Although these studies identified a parameter space 
 to realize the Higgs mass around 125 GeV, 
 phenomenology of the gravitino dark matter has not been studied in detail. 
In this paper, we complete our study on the mGMSB 
 by considering not only the realization of 125 GeV Higgs mass 
 but also the the gravitino dark matter phenomenology, 
 in particular, the so-called superWIMP scenario. 
We will show that once the values of $\tan \beta$ and 
 the number of messengers ($N_m$) are fixed, 
 the sparticle mass spectrum is completely determined 
 and the resultant sparticle mass spectrum is consistent  
 with all phenomenological constraints.

In the mGMSB, we introduce the superpotential 
 for the messenger sector, 
\bea 
 W_{mess} = \sum_{i=1}^{N_m} S \overline{\Phi}_i  \Phi_i ,    
\eea
 where $S$ is a chiral superfield in the hidden sector 
 with $ \langle S \rangle = M + \theta^2 F$, 
 and $\overline{\Phi}_i$ and $ \Phi_i $ are 
 a vector-like pair of messengers of 
 the $\bar{\bf 5}+ {\bf 5}$ representation  
 under the SU(5) GUT gauge group. 
For simplicity, we use the SU(5) GUT notation throughout the paper. 
In order to maintain the successful SM gauge coupling unification, 
 a pair of messengers should be in a complete SU(5) representation 
 and have an SU(5) invariant mass term.

Soft SUSY breaking terms can be extracted 
 from the SUSY wave function renormalization coefficients 
 with the threshold corrections by the messengers~\cite{GR}. 
The MSSM gaugino masses at a scale $\mu \leq M$ 
 is given by (we assume $ F \ll M^2$) 
\bea  
   M_i(\mu) = \frac{\alpha_i(\mu)}{4 \pi} \Lambda N_m,  
 \label{gauginomass}
\eea  
 where $i=1,2,3$ correspond to the SM gauge interactions 
 of SU(3), SU(2) and U(1)$_Y$, respectively, 
 and $\Lambda = F/M$ is the SUSY breaking parameter. 
When we neglect Yukawa coupling contributions, the MSSM scalar 
 squared masses at $\mu \leq M$ are given by 
\bea 
  m_{\tilde{f}}^2 (\mu) &=&
  \sum_i  2 C_i \left( \frac{\alpha_i(\mu)}{4 \pi} \right)^2
  \Lambda^2 \; N_m \; G_i(\mu, M) ,
 \label{scalarmass}
\eea
where $ G_i(\mu, M) = \xi_i^2 + \frac{N_m}{b_i} (1-\xi_i^2)$ 
 with $\xi_i  = \alpha_i(M)/\alpha_i(\mu)$. 
Here $b_i$ are the beta function coefficients for different 
 SM gauge groups, $C_i$ are the quadratic Casimir, 
 and the sum is taken corresponding to the representation
 of the sparticles under the SM gauge groups.
The messenger has no contribution to $A$-parameter 
 at the messenger scale, $A(M)=0$. 
In the following numerical analysis, we employ 
 the SOFTSUSY 3.3.1 package~\cite{softsusy} 
 to solve the MSSM RGEs and produce mass spectrum, 
 where the following free parameters are defined at the messenger scale:  
\bea 
 N_m, M, \Lambda, c_{\rm grav}, \tan \beta, {\rm sign}(\mu). 
\eea 
For simplicity, we set $\mu > 0$. 
In our analysis we assume $c_{\rm grav}=1$, 
 which means $F$ is the dominant SUSY breaking source. 
In this case, gravitino mass is given by 
\bea 
 m_{\tilde G} = \frac{M \Lambda}{\sqrt{3} M_P}, 
\eea
 where $M_P=2.4 \times 10^{18}$ GeV is the reduced Planck mass. 
Since flavor-dependent supergravity contributions to sfermion 
 masses are estimated as $\Delta m_{\tilde f}^2 \sim m_{\tilde G}^2$, 
 we need to set $ m_{\tilde G}^2 \ll m_{\tilde f}^2$, 
 equivalently, $M \ll 10^{16}$ GeV, 
 to make the flavor-independent GMSB contributions dominant. 
Therefore, gravitino is always the LSP and 
 hence the dark matter candidate. 
 From Eqs.~(\ref{gauginomass}) and (\ref{scalarmass}) 
 we see that the NLSP is most likely bino-like neutralino 
 for $N_m = {\cal O}(1)$.

As is well known, stop loop corrections play a crucial role 
 to push up the SM-like Higgs boson mass from its tree-level value, 
 $m_h \simeq m_Z \cos 2 \beta$. 
Since the corrections logarithmically depend on stop mass, 
 a large stop mass of ${\cal O}$(10 TeV) is necessary 
 to realize $m_h \simeq 125$ GeV. 
This corresponds to $\Lambda \sim 10^6$ GeV from Eq.~(\ref{scalarmass}) 
 and mass of the NLSP bino-like neutralino to be 
 $m_{\tilde B} \sim 1$ TeV from Eq.~(\ref{gauginomass}), 
 for $N_m = {\cal O}(1)$.

Now let us consider phenomenology for the gravitino dark matter. 
This is quite different from the usual one
 with the dark matter as a weakly interacting massive particle (WIMP), 
 because gravitino couples to the MSSM particles super-weakly and 
 (unless the gravitino is light, say $m_{\tilde G} \lesssim$ 1 keV) 
 it has never been in thermal equilibrium in history of the Universe. 
This fact brings uncertainty in evaluating the relic abundance 
 of the dark matter and thus one may think this scenario 
 less interesting. 
However, there is a very appealing possibility 
 with a super-weakly interacting dark matter particle, 
 the so-called superWIMP scenario~\cite{superWIMP1, superWIMP2} 
 (see also \cite{superWIMP3} for the superWIMP scenario 
  in non-standard cosmology). 
In this scenario, the superWIMP dark matter is mainly produced 
 via the decay of a long-lived WIMP, so that its relic abundance 
 is given as 
\bea 
 \Omega_{DM} h^2 = \Omega_{X} h^2 
  \times \left(\frac{m_{DM}}{m_X} \right), 
\eea 
where $\Omega_{X} h^2$ would be the thermal relic abundance of 
 the long-lived WIMP ($X$) if it were stable. 
In this scenario, nevertheless the superWIMP has 
 never been in thermal equilibrium, its relic abundance 
 is calculable as in the usual WIMP dark matter scenario.

We adopt the superWIMP scenario in the mGMSB, 
 where the superWIMP is the LSP gravitino 
 and the WIMP $X$ is the bino-like neutralino. 
Note that bino-like neutralino, if it were the LSP, 
 tends to be over-abundant because of its small annihilation 
 cross section. 
Therefore, the suppression factor $m_{\tilde{G}}/m_{\tilde B} \ll 1$ 
 can work well to adjust $\Omega_{DM} h^2$ 
 to be the observed value~\cite{goldilocks}. 
With a fixed $\tan \beta$ and $N_m$, 
 we first calculate particle mass spectrum by SOFTSUSY 3.3.1 package 
 for various values of $M$ and $\Lambda$. 
Then, the relic abundance of the NLSP neutralino is calculated 
 by using the micrOMEGAs 2.4.5~\cite{micrOMEGAs} 
 with the output of SOFTSUSY in the SLHA format~\cite{SLHA}. 
Multiplying by the factor $m_{\tilde{G}}/m_{\tilde B}$, 
 we finally obtain the relic abundance of the gravitino dark matter.

\begin{figure}[htbp]
\begin{center}
{\includegraphics[width=1.0\columnwidth]{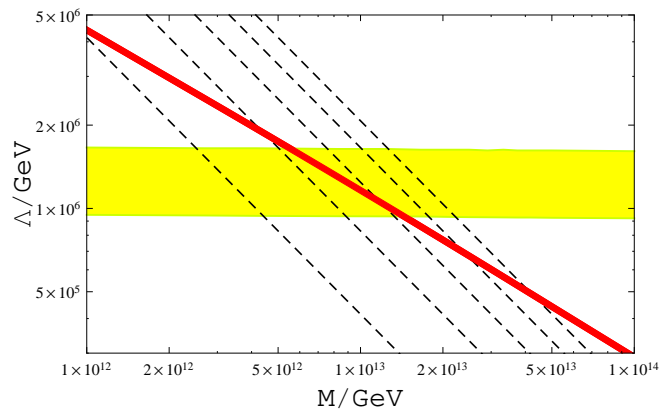}}
\caption{
Contours on the ($M, \Lambda$)-plane  
 for $\tan \beta =10 $ and $N_m=1$. 
The dark shaded region (in red) 
 and the horizontal shaded region (in yellow) 
 satisfy the constraints of 
 the observed dark matter abundance
 ($0.1064 \leq \Omega_{\tilde G} h^2 \leq 0.1176$) 
 and the SM-like Higgs boson mass 
 (124 GeV $\leq m_h \leq$ 126 GeV), 
 respectively. 
}
\label{Fig1}
\end{center}
\end{figure}

Fig.~1 shows our numerical results in the ($M, \Lambda$)-plane, 
 for $\tan \beta =10 $ and $N_m=1$. 
In the dark shaded region (in red), the resultant gravitino
 relic abundance is consistent with the observed value~\cite{WMAP}, 
 $\Omega_{\tilde G} h^2 = 0.1120 \pm 0.0056$. 
The parameter sets in the horizontal shaded region (in yellow) 
 predict the SM-like Higgs boson mass in the range of 
 124 GeV $\leq m_h \leq$ 126 GeV. 
The dashed lines correspond to the gravitino mass 
 $m_{\tilde G} =1,2,3,4,$ and $5$ GeV, respectively, from left to right. 
The parameter region which simultaneously satisfies 
 the observed Higgs boson mass and the dark matter relic abundance 
 can be pined down by the overlap of the two shaded regions, 
 $(M, \Lambda) \simeq 
 (9.2 \times 10^{12}~{\rm GeV}, 1.2 \times 10^6~{\rm GeV})$.

Here we give a semi-analytical explanation of our results. 
The relic abundance of the bino-like neutralino is well approximated
 as~\cite{ADG}
\bea 
 \Omega_{\tilde B} h^2 \simeq 
 \frac{8.7 \times 10^{-11} \; {\rm GeV}^{-2} \; (n+1) x_f^{n+1}}
{\sqrt{g_*} \sigma_{\tilde B}}, 
\eea   
 where $n=1$, $x_f = m_{\tilde B}/T_f \sim 20$ is the freeze-out temperature, 
 and $ \sigma_{\tilde B} $ is the pair annihilation cross section 
 of the bino-like neutralino through the exchange of 
 the right-handed charged sleptons in the $t$-channel, 
\bea 
 \sigma_{\tilde B} \simeq 
 \frac{3 g^4 \tan^4 \theta_W r (1+r^2) }
 {2 \pi m_{{\tilde e}_R}^2 (1+r^4)}
\eea
 with $r = (M_1/m_{{\tilde e}_R})^2$. 
Using Eqs.~(\ref{gauginomass}) and (\ref{scalarmass}), 
 we can check that the factor $r (1+r^2)/(1+r^4)$ 
 is almost constant $\sim 0.12$ 
 for various values of $N_m={\cal O}(1)$ and $M=10^{12-14}$ GeV. 
Thus, we find 
\bea 
 \Omega_{\tilde G} h^2 \simeq 
 \Omega_{\tilde B} h^2 \times \left( \frac{m_{\tilde G}}{M_1} \right)
 \propto  m_{{\tilde e}_R}^2 \frac{m_{\tilde G}}{M_1} 
 \propto M \Lambda^2.  
\eea
Since sfermion masses logarithmically depend on the messenger
 scale (see Eq.~(\ref{scalarmass})), 
 the predicted Higgs mass determined by stop masses 
 is almost independent of $M=10^{12-14}$ GeV.

\begin{figure}[htbp]
\begin{center}
{\includegraphics[width=1.0\columnwidth]{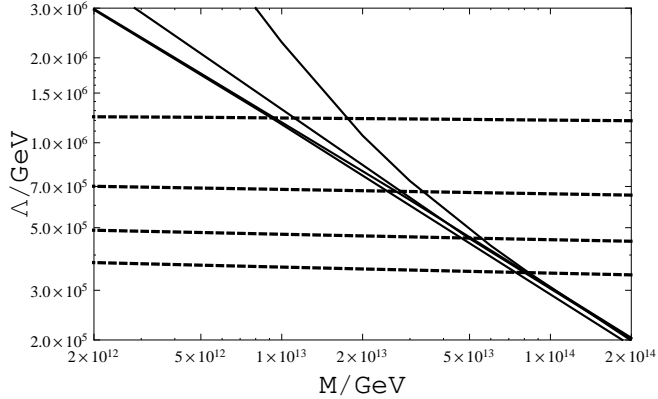}
\caption{
Contours on the ($M, \Lambda$)-plane  
 for $\tan \beta =10$ and $N_m=1,2,3,4$. 
The solid lines correspond to the region satisfying 
 $\Omega_{\tilde G} h^2 =0.112$, from left to right 
 for $N_m =1,2,3,4$, respectively. 
Along the dashed lines 
 the SM-like Higgs boson mass is predicted to be $m_h=125$ GeV, 
 from top to bottom for $N_m =1,2,3,4$, respectively. 
}
\label{Fig2}}
\end{center}
\end{figure}

We repeat the similar numerical analysis 
 for various values of $N_m$ and $\tan \beta$. 
Fig.~2 depicts the results for $N_m=1,2,3,4$, 
 with $\tan \beta =10$. 
The solid lines show the contours along which 
 $\Omega_{\tilde G} h^2 = 0.112$ is satisfied,
 from left to right corresponding to $N_m=1,2,3,4$, respectively. 
All lines are very close to each other except that 
 the line for $N_m=4$ is substantially deviating from 
 the other lines for $M \lesssim 5 \times 10^{13}$ GeV. 
This is because for such a parameter region, 
 the NLSP bino-like neutralino becomes more degenerate 
 with lighter stau as $M$ is lowered, 
 so that the co-annihilation process of the NLSP with the stau 
 becomes more effective and reduces the abundance. 
In order to compensate this reduction and achieve 
 the observed relic abundance, 
 a relatively higher neutralino mass, 
 equivalently a larger $\Lambda$ is needed. 
The dashed horizontal lines correspond to 
 the Higgs boson mass prediction $m_H=125$ GeV, 
 from top to bottom for $N_m=1,2,3,4$, respectively. 
We find that stau becomes the NLSP for $N_m \geq 5$. 
The annihilation process of the NLSP stau 
 is very efficient, so that the suppression factor 
 $m_{\tilde G}/m_X$ makes the gravitino abundance too small. 
Hence, we do not consider the case with $N_m \geq 5$.

In Fig.~2, we can pin down the parameter sets $(M, \Lambda)$ 
 which simultaneously satisfy the two conditions, 
 $\Omega_{\tilde G} h^2 = 0.112$ and $m_h=125$ GeV, 
 at the intersections of the solid and dashed lines, 
 for each $N_m$ value. 
In Table~1, we list particle mass spectra for $N_m=1,2,4$. 
Here we also list the result for $N_m=1$ and $\tan \beta =45$ 
 as another sample.

\begin{table}
\begin{center}
\begin{math}
\begin{array}{|c|c|c|c||c|}
\hline  
\tan \beta & \multicolumn{3}{|c|}{10} & 45 \\ 
\hline
N_m & 1 & 2  & 4 & 1 \\
\hline
M    & 9.16 \times 10^{12} & 2.66 \times 10^{13}
     & 8.28 \times 10^{13} 
     & 1.51 \times 10^{13} \\
\Lambda & 1.23 \times 10^6 & 6.73 \times 10^5 
        & 3.47 \times 10^5 
        & 1.05 \times 10^6 \\
\hline
  h_0 & \multicolumn{4}{|c|}{125} \\ 
\hline
  H_0 & 7123 & 6304 & 5607 &  4418  \\ 
  A_0 & 7123 & 6304 & 5607 &  4419  \\
H^\pm & 7123 & 6304 & 5608 &  4419  \\  
\hline 
\tilde{g} & 7726 & 8300 & 8424 & 6719  \\ 
{\tilde{\chi}^0}_{1,2} 
  & 1693, 3141 & 1856, 3424 & 1909, 3509 & 1454, 2707 \\ 
{\tilde{\chi}^0}_{3,4} 
  & 5147, 5148 & 4717, 4719 & 4365, 4369 & 4370, 4372 \\ 
 {\tilde{\chi}^{\pm}}_{1,2} 
  & 3141, 5149 & 3424, 4719 & 3509, 4369 & 2707, 4372 \\ 
\hline 
\tilde{u},\tilde{c}_{1,2} 
 & 9227, 10123 & 8491, 9193 & 7837, 8389 & 7983, 8757 \\
\tilde{t}_{1,2} 
 & 7274, 9260  & 6829, 8455 & 6413, 7745 & 6296, 7695 \\
\tilde{d},\tilde{s}_{1,2} 
 & 9034, 10123 & 8346, 9193 & 7733, 8389 & 7812, 8757 \\
\tilde{b}_{1, 2}  
 & 8990, 9258 & 8308, 8452  & 7698, 7742 & 7107, 7693 \\ 
\hline 
\tilde{\nu}_{e, \mu}  
 & 4989 & 4242 & 3580 & 4330 \\
\tilde{\nu}_{\tau} 
 & 4976 & 4231 & 3571 & 4084 \\
\tilde{e},\tilde{\mu}_{1,2} 
 & 3305, 4990  & 2783, 4243 & 2290, 3581 & 2895, 4330 \\
\tilde{\tau}_{1,2}  
 & 3267, 4977  & 2750, 4232 & 2262, 3572 & 2070, 4087 \\
\hline 
m_{\tilde G} & 2.66 & 4.24 & 6.81 & 3.77 \\
\hline \Omega_{\tilde G} h^2 
 & \multicolumn{4}{|c|}{0.112} \\
\hline 
\end{array}
\end{math}
\caption{
Particle mass spectra (in GeV) for various $N_m$ and $\tan \beta$. 
The fundamental parameters in the mGMSB, 
 $M$ and $\Lambda$, have been fixed 
 so as to satisfy two independent conditions, 
 $\Omega_{\tilde G} h^2=0.112$ and $m_h=125$ GeV. 
} 
\end{center}
\end{table}

In the superWIMP scenario, the NLSP has a long lifetime 
 and its late-time decay is potentially dangerous 
 for the success of big bang nucleosynthesis (BBN). 
In our scenario, the NLSP is bino-like neutralino 
 and its lifetime is estimated as~\cite{superWIMP1} 
\bea 
 \tau_{\tilde B} \simeq 0.74 \; {\rm sec} \times 
  \left( \frac{m_{\tilde G}}{1 \; {\rm GeV}} \right)^2 
  \left( \frac{1 \; {\rm TeV}}{m_{\tilde B}} \right)^5 
\eea 
 for its main decay mode ${\tilde B} \to \gamma + {\tilde G}$. 
For the parameter sets we have determined, 
 the formula leads to $\tau_{\tilde B} < 1$ sec. 
Therefore, the NLSP neutralinos decay before the BBN era 
 $\sim 1$ sec and the success of BBN remains intact.

The gravitino dark matter is produced by the decay of 
 the bino-like neutralino and carries a large kinetic energy 
 when it is produced. 
Such a non-thermally produced dark matter behaves 
 as a ``warm'' dark matter, rather than the cold dark matter (CDM). 
While the CDM scenario has made great success 
 in the structure formation of the Universe at large scales 
 $> 1$ Mpc, the recent high resolution N-body simulations 
 showed that the CDM scenario predicts too much power 
 on sub-galactic scales to be consistent with the observations. 
It has been pointed out~\cite{LHZB} that this problem can 
 be solved by a non-thermally produced dark matter 
 if its comoving free-streaming scale ($\lambda_{\rm FS}$)
 at the time of the matter-radiation equality 
 ($t_{\rm EQ} \simeq 2.0 \times 10^{12}$ sec) is around $0.1$ Mpc.

The free-streaming scale of the gravitino dark matter
 can be calculated as~\cite{FS}
\bea 
& \lambda_{\rm FS}& 
 = \int_{\tau_{\tilde B}}^{t_{\rm EQ}} \frac{v(t')}{a(t')} dt 
 \simeq 2 v_0 t_{\rm EQ} (1 + z_{\rm EQ})^2 \nonumber \\
&\times& 
 \log \left[ 
 \sqrt{1+\frac{1}{v_0^2 (1+z_{\rm EQ})^2 }}+ \frac{1}{v_0 (1+z_{\rm EQ})}
 \right], 
\eea
 where $z_{\rm EQ} \simeq 3000$ is the red shift at $t_{\rm EQ}$, and 
 $ v_0 \simeq \frac{T_0}{T_I} \frac{E_I}{m_{\tilde G}}$ 
 is the current velocity of the gravitino dark matter 
 with the present temperature of the cosmic microwave background $T_0$, 
 the temperature of the Universe 
 $T_I \simeq \sqrt{1 \; {\rm sec}/\tau_{\tilde B}} \times 10^{-3}$ GeV, 
 and the energy of the gravitino $E_I = m_{\tilde B}/2$ 
 when the gravitino is produced from the NLSP neutralino decay. 
It is easy to find 
 $ v_0 \simeq 1.0 \times 10^{-7} 
 \left( \frac{1 \; {\rm TeV}}{m_{\tilde B}} \right)^{3/2}$, 
 which is independent of the gravitino mass,  
 and the free-streaming scale is 
\bea 
 \lambda_{\rm FS} \simeq 0.18 \; {\rm Mpc} \times 
 \left( \frac{1 \; {\rm TeV}}{m_{\tilde B}} \right)^{3/2}.  
\eea
The mass spectra listed on Table~1 give the free-streaming scale 
 $ \lambda_{\rm FS} \simeq 0.1$ Mpc,
 which is nothing but the value suitable for solving 
 the problem with the CDM scenario on sub-galactic scales.

\begin{center}
{\bf Acknowledgments}
\end{center}
The author would like to thank Hieu Minh Tran 
 and Osamu Seto for reading the manuscript 
 and useful comments. 
This work is supported in part by the DOE 
 under grant No. DE-FG02-10ER41714.

%


\end{document}